\documentclass[openacc]{rsproca_new}%%%%where rsproca is the template name

\usepackage{blindtext}
\usepackage{mathrsfs}
\PassOptionsToPackage{square,comma,numbers,sort&compress}{natbib}
\usepackage{natbib}
\usepackage{amsmath,bm}
\usepackage{mathtools}
\usepackage{orcidlink}
\usepackage[dvipsnames]{xcolor}
\usepackage{labelfig}
\usepackage{bm}
\usepackage{hyperref}
\usepackage{lineno}
\usepackage[dvipsnames]{xcolor}

\titlehead{Research}

\begin{document}

%%%% Article title to be placed here
\title{Shear-driven mixing of segregated granular materials}

\author{%%%% Author details
Hugo N. Ulloa$^{1}$\orcidlink{0000-0002-1995-6630} and Tomás Trewhela$^{2}$\orcidlink{0000-0002-7461-8570}}

%%%%%%%%% Insert author address here
\address{$^{1}$Department of Earth and Environmental Science, University of Pennsylvania, Philadelphia, PA19104, USA\\
$^{2}$Facultad de Ingeniería y Ciencias,
Universidad Adolfo Ibañez, Av. Padre Hurtado 750, 2562340 Viña del Mar, Valparaíso, Chile}

%%%% Subject entries to be placed here %%%%
\subject{soft earth geophysics, granular matter, geophysical flows}

%%%% Keyword entries to be placed here %%%%
\keywords{granular flows, particle-size segregation, bidisperse mixing length, degree of mixing}

%%%% Insert corresponding author and its email address}
\corres{H.N.~Ulloa\\ \email{ulloa@sas.upenn.edu}\\ T.~Trewhela\\\email{tomas.trewhela@uai.cl}}

%%%% Abstract text to be placed here %%%%%%%%%%%%
\begin{abstract}
As granular materials flow and settle, interactions among particles of different sizes or properties drive mixing and segregation, producing dynamics that shape systems ranging from silos to asteroids. A hallmark of polydisperse granular flows is shear-driven size segregation, in which larger grains tend to rise above smaller ones through particle-scale rearrangements. Despite substantial progress in modeling granular flow and segregation, the complementary process of granular mixing remains less well understood. Here, we investigate the evolution of initially segregated dense granular materials driven out of equilibrium by imposed shear. We ask: what controls the extent and rate of mixing and restratification in a sheared bidisperse granular flow? Addressing this question is essential for understanding how external forcing disrupts or reinforces particle-size organization, and for optimizing processes that require controlled mixing. Using theoretical analysis and numerical simulations, we develop a framework that quantifies the degree of mixing and segregation dynamics of dense bidisperse granular flows. Our results identify the controlling roles of particle-size ratio and the Péclet number, clarify the conditions under which segregated states persist or are transiently homogenized, and provide a basis for improved prediction and control of granular mixtures in natural and industrial settings.
\end{abstract}
%%%%%%%%%%%%%%%%%%%%%%%%%%%

%\rsbreak

%%%%%%%%%% Insert the texts which can accomdate on firstpage in the tag "fmtext" %%%%%
\maketitle
\section{Introduction}

\begin{figure}[!h]
\centering\includegraphics[width=\textwidth]{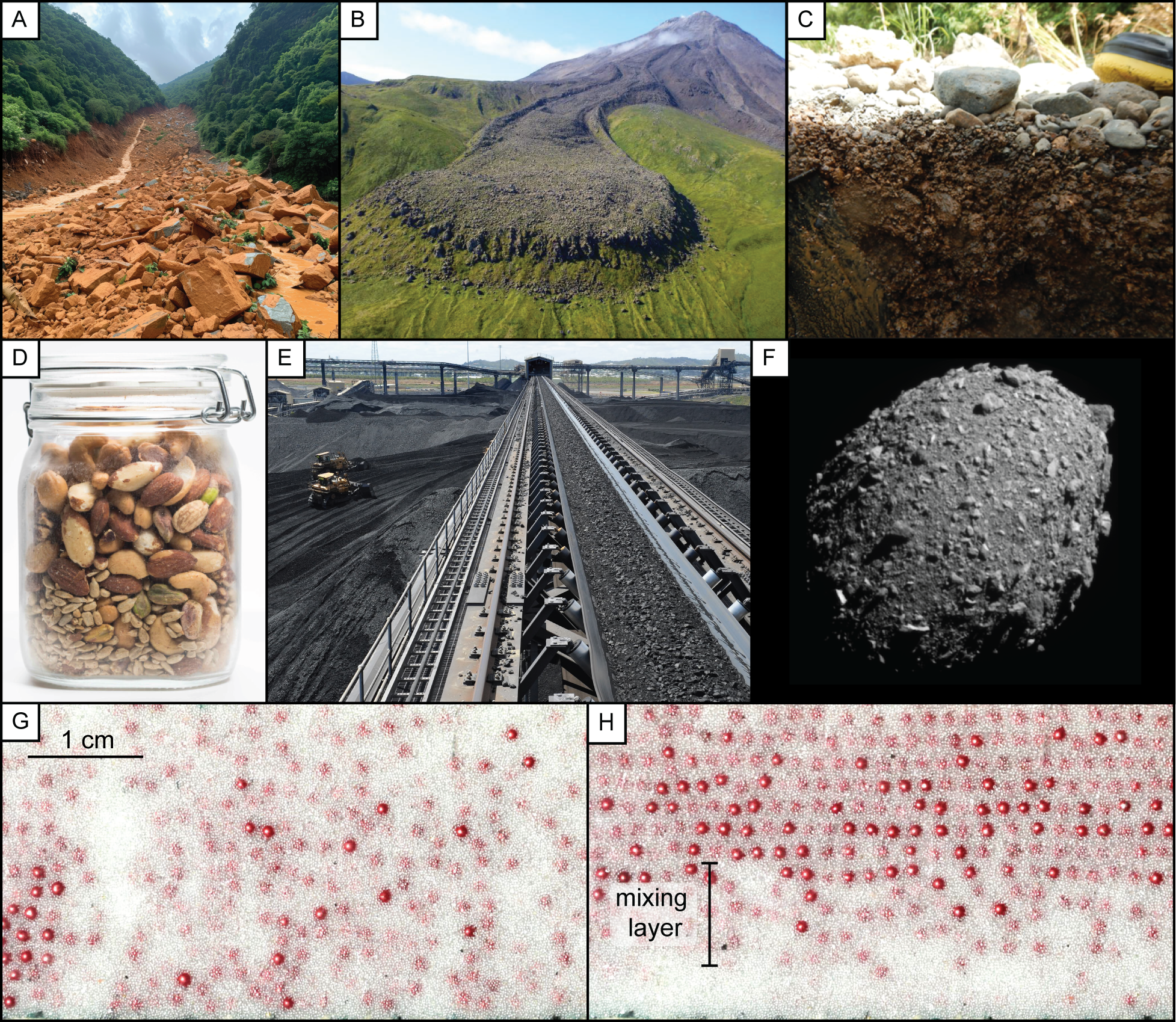}
%%% where xxxxxx name represents "figurename.eps"
\caption{Granular systems experiencing particle size segregation: (A) Debris flow \cite{zhao2025scaling}; (B) Kanaga Volcano with its segregated lava flow deposit from the 1906 eruption. [Credit: Coombs, M. L., Alaska Volcano Observatory / U.S. Geological Survey]; (C) Riverbed armoring \textcolor{black}{in Rio de la Plata [Credit: Andrew Branard]}; (D) Brazilian nut effect [credit: J. Adam Fenster]; (E) Mined coal mobilized by conveyor belt [Credit: Dave Hunt/AAP]; (F) \textcolor{black}{Pre-impact image of the rubble-pile asteroid Dimorphos and its segregated surface captured by the DART mission [Credit: NASA/Johns Hopkins APL]. Experiment illustrating disordered mixed state (G) and partially segregated granular flow (H) [Credit: Chris Johnson] \citep{gamble2026reversed}}.}
\label{fig:1}
\end{figure}

% opening
Polydisperse granular flows (\textcolor{black}{figure~\ref{fig:1}}) are governed by the competition between two antagonistic transport mechanisms: size segregation, which sharpens concentration gradients through size-selective particle motion, and diffusive remixing, which smooths these gradients through random particle rearrangements \cite{Ottino2000,Gray2018}. In dense sheared granular media, this competition determines whether an initially stratified configuration remains stable, is transiently eroded into a mixed state, or relaxes back towards a normally graded segregated state. It also sets the characteristic thickness of the polydisperse mixing-layer that develops between segregated regions (\textcolor{black}{figure~\ref{fig:1}H)}. However, a predictive description of how this balance controls the relevant timescales and characteristic length scales remains incomplete.

% context
This problem arises generically in granular materials, whose discrete and frictional nature gives rise to strong coupling between particle rearrangement, frictional contacts, stresses, and bulk deformation \citep{Ottino2000,Gray2018,Breard2019,Zhou2020}. In polydisperse systems, shear commonly drives size segregation, with large grains tending to rise while small grains percolate downwards \citep{Middleton70,SavageLun1988}. The resulting spatial sorting modifies the internal structure, transport properties, and effective rheology of the material, with important consequences across geophysical and industrial settings \citep{Zhou2020,Trewhela2021a,Trewhela2021b,russell2023plastic,Cunez2024,Trewhela24,TrewhelaUlloa2024,trottet2025sandball}.

% previous work and open gaps
Substantial progress has been made in understanding segregation in dense granular flows. Mainstream mechanisms such as kinetic sieving and squeeze expulsion, together with continuum convection--segregation--diffusion equations, have clarified how shear, pressure and particle-size contrasts generate net segregation fluxes and stratified states \citep{SavageLun1988,GrayThornton2005,ThorntonGrayHogg2006,Schlick2016,Trewhela2021a,jing2021unified,navarrete2026stress}. More recent studies have quantified segregation rates, characterised intruder dynamics, and developed continuum descriptions that link segregation to rheology and the mechanical energy budget of the flow \citep{Trewhela2021a,Trewhela2021b,TrewhelaUlloa2024,zhao2025scaling}. By comparison, the transient dynamics of diffusive remixing remain much less well resolved. In most continuum models, remixing enters as a diffusive counter-flux opposing segregation \citep{Ottino2000,Schlick2016,Fry2019,Golick09,May2010,Trewhela24,Singh24}, but key questions remain open: the accessible range of mixing and segregation in sheared systems, the thickness of the coexistence region produced by the balance between segregation and diffusion, and the times required to reach the maximally mixed and asymptotically segregated states.

These issues are especially relevant for initially segregated beds subjected to imposed shear. Here, the flow is driven away from equilibrium, so that segregation no longer simply builds structure from a mixed state, but instead interacts with diffusive remixing to breakdown and subsequently rebuild stratification. This competition is expected to be governed primarily by two parameters: the particle-size ratio $R_d$, which sets the strength of size-selective percolation, and the Péclet number $Pe$, which measures the relative importance of segregation and remixing \cite{Wiederseiner2011,Fan2014,Trewhela2021a}. The problem is, therefore, to determine how these parameters control the destruction, persistence, and recovery of stratification. In particular, we raise the following questions:
\vspace{-0.25cm}
\begin{itemize}
\item[(i)] Does the degree of segregation systematically increase with increasing $R_d$ and $Pe$?
\item[(ii)] Does the polydisperse mixing-layer thickness decrease systematically as segregation becomes stronger relative to remixing?
\item[(iii)] Do the timescales for reaching maximum mixing and recovering the stable segregated state follow distinct scalings with $R_d$ and $Pe$?
\end{itemize}

\vspace{-0.25cm}
% This paper paragraph
Here, we address these questions by studying the shear-driven evolution of an initially segregated bidisperse granular column across the parameter space $(R_d,Pe)$.
\textcolor{black}{Our analysis builds on continuum theory for size-driven segregation and diffusive remixing in dense granular flows \citep{Gray2018,thornton2026modeling}, together with a numerical simulations for resolving the concentration dynamics (Section~\ref{sec:2}).}
\textcolor{black}{We use this framework to derive a scalar measure} of segregation and examine its transient and steady-state dynamics as functions of $R_d$ and $Pe$ \textcolor{black}{(Section~\ref{sec:3})}. 
We then analyze the bidisperse mixing-layer that \textcolor{black}{forms between large- and small-particle-rich regions, showing how its thickness is controlled by the local balance between segregation fluxes and diffusive remixing} \textcolor{black}{(Section~\ref{sec:4})}. \textcolor{black}{Finally}, we identify two characteristic timescales: a mixing timescale associated with the breakdown of segregation and the attainment of maximum mixing, and a segregation timescale associated with restratification toward the asymptotically stable stratified state \textcolor{black}{(Section~\ref{sec:5})}. Taken together, our results provide a tractable analytical description of how transient mixing and restratification depend on $R_d$ and $Pe$, clarify the conditions under which segregated states persist, are transiently disrupted, or become efficiently homogenized, \textcolor{black}{and motivate future research directions (Section~\ref{sec:6})}.

\section{Framework}\label{sec:2}

\subsection{Convective-diffusive equation for segregation}

We consider a non-cohesive bidisperse granular medium, composed of rigid spheres of diameters $d_s$ and $d_l(>d_s)$, and uniform density $\rho_*$, occupying a fixed volume $V$. \textcolor{black}{The solid volume in the system is $V_{\rm solid}=V_s+V_l$, where $V_s=n_s \frac{\pi}{6}d_s^3$ and $V_l=n_l \frac{\pi}{6}d_l^3$}, for $n_s$ small and $n_l$ large particles. \textcolor{black}{The global solid volume fraction is therefore $\Phi=(V_s+V_l)/V=\Phi_s+\Phi_l$.}

In dense granular flows, $\Phi$ is typically nearly constant \cite[e.g.,][]{gdr2004dense,houssais2015onset,Trewhela2021b}, so we introduce the species concentrations $\phi_s=\Phi_s/{\Phi}$, $\phi_l=\Phi_l/{\Phi}$, which satisfy 
 $\phi_s+\phi_l=1$. Treating the mixture as a continuum, the transport of each species is governed by a segregation--diffusion balance,
\begin{equation}\label{eq:phi_nu}
    \frac{\partial \Phi_\nu}{\partial t}
    +\nabla\cdot \bm{F}_{\Phi_\nu}
    =
    \nabla\cdot\left(\mathcal{D}_{sl}\nabla \Phi_\nu\right),
\end{equation}
for $\nu\in\left\{s,l\right\}$, where $\Phi_\nu(t,z)$ denotes the partial solid volume fraction of species $\nu$ as a function of time `t' and height `z', $\bm{F}_{\Phi\nu}$ is the segregation flux, and $\mathcal{D}_{sl}$ is the bidisperse diffusivity. The segregation flux is written as
\begin{equation}
    \bm{F}_{\Phi_\nu}=
    \begin{cases}
    \displaystyle
    +\,f_{sl}\,\Phi_s(1-\phi_s)\,\dfrac{\bm{g}}{|\bm{g}|}, & \nu=s,\\[8pt]
    \displaystyle
    -\,f_{sl}\,\Phi_l(1-\phi_l)\,\dfrac{\bm{g}}{|\bm{g}|}, & \nu=l,
    \end{cases}
\end{equation}
where $\bm{g}$ is the gravitational acceleration. This form expresses the net segregation transport along the gravity direction, with opposite fluxes for the two species. Following the scaling arguments and experimental results by Trewhela~et~al.~\cite{Trewhela2021a}, the segregation velocity scale $f_{sl}$ is modelled as
\begin{equation}
    f_{sl}
    =
    \left(
    \frac{\mathcal{B}\rho_* g \dot{\gamma}\bar{d}^{\,2}}{p}
    \right)
    \left[
    (R_d-1)+\mathcal{E}(1-\phi_s)(R_d-1)^2
    \right],
\end{equation}
where $\mathcal{B}=0.374$ and $\mathcal{E}=2.096$ are empirical coefficients \cite{Trewhela2021a}, $R_d=d_l/d_s$ is the particle-size ratio, $p$ is the local pressure, and $\dot{\gamma}$ is the shear rate. The concentration-dependent mean particle diameter is
$\bar{d}=\phi_s d_s+\phi_l d_l=(1-\chi_d\phi_s)d_l$, with $\chi_d=(R_d-1)/R_d$ the size-asymmetry coefficient \citep{Gajjar14,vanderVaart15,Trewhela24}. This form of $f_{sl}$ captures the experimentally observed dependence of segregation on shear rate, pressure, local composition, and particle-size contrast \citep{Golick09,Thornton12,Tripathi21}.

The diffusive remixing is represented through the constitutive relation $\mathcal{D}_{sl}=\mathcal{A}\dot{\gamma}\bar{d}^{\,2}$, where $\mathcal{A}=0.108$ \citep{Utter04}. Physically, this term accounts for collisional diffusion arising from shear-induced particle fluctuations, which opposes segregation.

Together, segregation and diffusion balance out in the form of the segregation Péclet number

\begin{equation}
    Pe=\frac{hf_{sl}}{\mathcal{D}_{sl}}=\frac{\mathcal{B}\rho_{*}gh}{\mathcal{A}\,p}\left[(R_{d}-1)+\mathcal{E}(1-\phi_{s})(R_{d}-1)^{2}\right],
    \label{eq:Peclet}
\end{equation}
Here, the Péclet number controls the segregation-mixing dynamics and ultimately determines the long-term degree of mixing.

\subsection{Simulations}

We investigate segregation and mixing dynamics governed by the equation~\eqref{eq:phi_nu}.
\textcolor{black}{For this system, we impose adiabatic boundaries at the base ($z=0$) and upper surface ($z=h$) of the granular layer, so that no particles leave or enter the domain. Since the governing equation for each species $\nu$ is written in conservative form, the appropriate boundary condition is zero total vertical species flux, rather than zero concentration gradient alone. Thus, for each species $\nu$, we impose
$$\mathcal{J}_{\nu}(t,z)=0,\quad \textrm{at}\quad z=0,\,h,$$
where $\mathcal{J}_{\nu}$ denotes the sum of the segregation and diffusive remixing fluxes appearing in right-hand of equation~\eqref{eq:phi_nu}. The initial condition is a perfectly segregated state, with small particles occupying the either the upper half of the layer and large particles occupying the lower half,
$$\phi_s(t,z)=\mathcal{H}(z-h/2), \qquad \phi_l(t,z)=1-\phi_s(t,z) \quad \textrm{at}\quad t=0,$$
or the opposite,
$$\phi_s(t,z)=1-\mathcal{H}(z-h/2), \qquad \phi_l(t,z)=1-\phi_s(t,z) \quad \textrm{at}\quad t=0.$$}

\textcolor{black}{The entire bidisperse granular layer is subjected to a uniform shear rate $\dot{\gamma}$, imposed by translating the upper boundary $(z=h)$ horizontally at speed $u_0$ while keeping the lower boundary $(z=0)$ fixed.}

We numerically integrated the coupled nonlinear PDE system in equation~\eqref{eq:phi_nu}, \textcolor{black}{subject to the boundary conditions described above}, using the method of lines (MOL) \cite{Schiesser2012}. \textcolor{black}{In this approach, the vertical coordinate is discretized into $n_z=250$ grid points, while time is kept continuous. The scalar field $\Phi_\nu(t,z)$ is therefore represented by its nodal values $\Phi_\nu(t,z_i)$, and the spatial derivatives in the governing equations are approximated by finite differences. This converts the PDE system into a coupled system of ordinary differential equations for the temporal evolution of the concentration at each vertical location. The term MOL reflects this construction: each fixed grid point $z_i$ defines a line parallel to the time axis in the $(z,t)$ plane, along which the local value of $\Phi_\nu$ evolves. The resulting discrete system was advanced in time using an adaptive RK45 time integrator.}

\begin{table}
  \begin{center}
  \vspace{0.2cm}
  $\mu_{1}=0.342$,\quad $\mu_{2}=0.557$,\quad $I_{0}=0.069$, \quad $\mu_{w}=0.8$, \quad $p_0=1000$ Pa, \quad $u_0=1$ m/s  \\
  \quad $d_l=143$ $\mu$m, \quad $\Phi=0.6$, \quad $\rho_{*}=2500$ kg/m$^3$, \quad $\mathcal{A}=0.108$, \quad $\mathcal{B}=0.374$, \quad $\mathcal{E}=2.096$
  \vspace{0.2cm}
  \caption{Parameters for numerical simulations. Frictional parameters  $\mu_{1}$, $\mu_{2}$ and $I_{0}$ of the $\mu(I)$-rheology measured by \cite{Barker17} for $d_{l}=$143 µm glass beads, \textcolor{black}{with $I_{0}$ a fitting parameter for the rheology}. The small particle diameter $d_s$ is determined by $d_s=R_{d}d_l$, with $R_d$ \textcolor{black}{the particle-size ratio} varying in each numerical solution. Boundary condition parameters at the wall: friction coefficient $\mu_w$, pressure $p_{0}$ and velocity $u_{0}$. Grains' intrinsic density $\rho_{*}$, solids volume fraction $\Phi$, and the universal parameters for  diffusivity $\mathcal{A}$ \citep{Utter04} and segregation $\mathcal{B}$, $\mathcal{E}$ \citep{Trewhela2021a}.}
  \label{tab:params}
  \end{center}
\end{table}

The pressure was prescribed as spatially uniform, $p=p_{0}$, with a constant shear rate $\dot{\gamma}=u_{0}/h$. To connect the model with dense granular flow rheology, $h$ was calculated using the $\mu(I)$-rheology parameters for the material \cite{Trewhela24}, $$h=\left(\frac{\mu_{w}}{\mu_1}-1\right)\frac{p_{0}}{\rho_{*}\,g\,\Phi}.$$
The parameter values used to calculate $h$, along with other particle parameters, were originally taken from \cite{Barker17}, which are detailed in table~\ref{tab:params}.

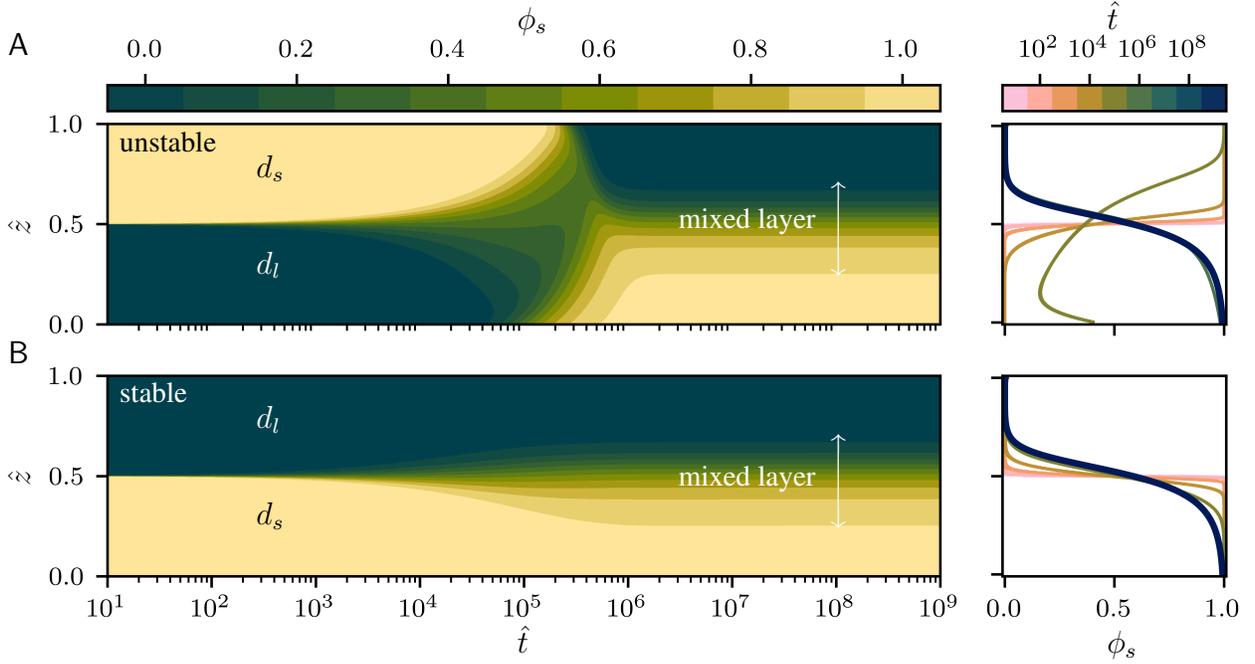
\begin{figure}[!h]
\begin{center}
\SetLabels
\L (0.0*0.93) $\sf A$\\
\L (0.0*0.45) $\sf B$\\
%\L (0.76*0.93) ($c$)\\
%\L (0.76*0.45) ($d$)\\
\L (0.0*0.65) \rotatebox{90}{$\hat{z}$}\\
\L (0.0*0.26) \rotatebox{90}{$\hat{z}$}\\
\L (0.41*0.0) $\hat{t}$\\
\L (0.885*0.0) $\phi_s$\\
\L (0.09*0.78) \textcolor{black}{\sf unstable}\\
\L (0.09*0.39) \textcolor{white}{\sf stable}\\
\L (0.55*0.66) \textcolor{white}{\sf mixed layer}\\
\L (0.55*0.26) \textcolor{white}{\sf mixed layer}\\
\L (0.66*0.647) \textcolor{white}{$\Bigg\updownarrow$}\\
\L (0.66*0.255) \textcolor{white}{$\Bigg\updownarrow$}\\
\L (0.2*0.74) \textcolor{black}{$d_{s}$}\\
\L (0.2*0.59) \textcolor{white}{$d_{l}$}\\
\L (0.2*0.35) \textcolor{white}{$d_{l}$}\\
\L (0.2*0.20) \textcolor{black}{$d_{s}$}\\
\L (0.41*0.97) $\phi_{s}$\\
\L (0.885*0.97) $\hat{t}$\\
\endSetLabels
\strut\AffixLabels{\includegraphics[width=\textwidth]{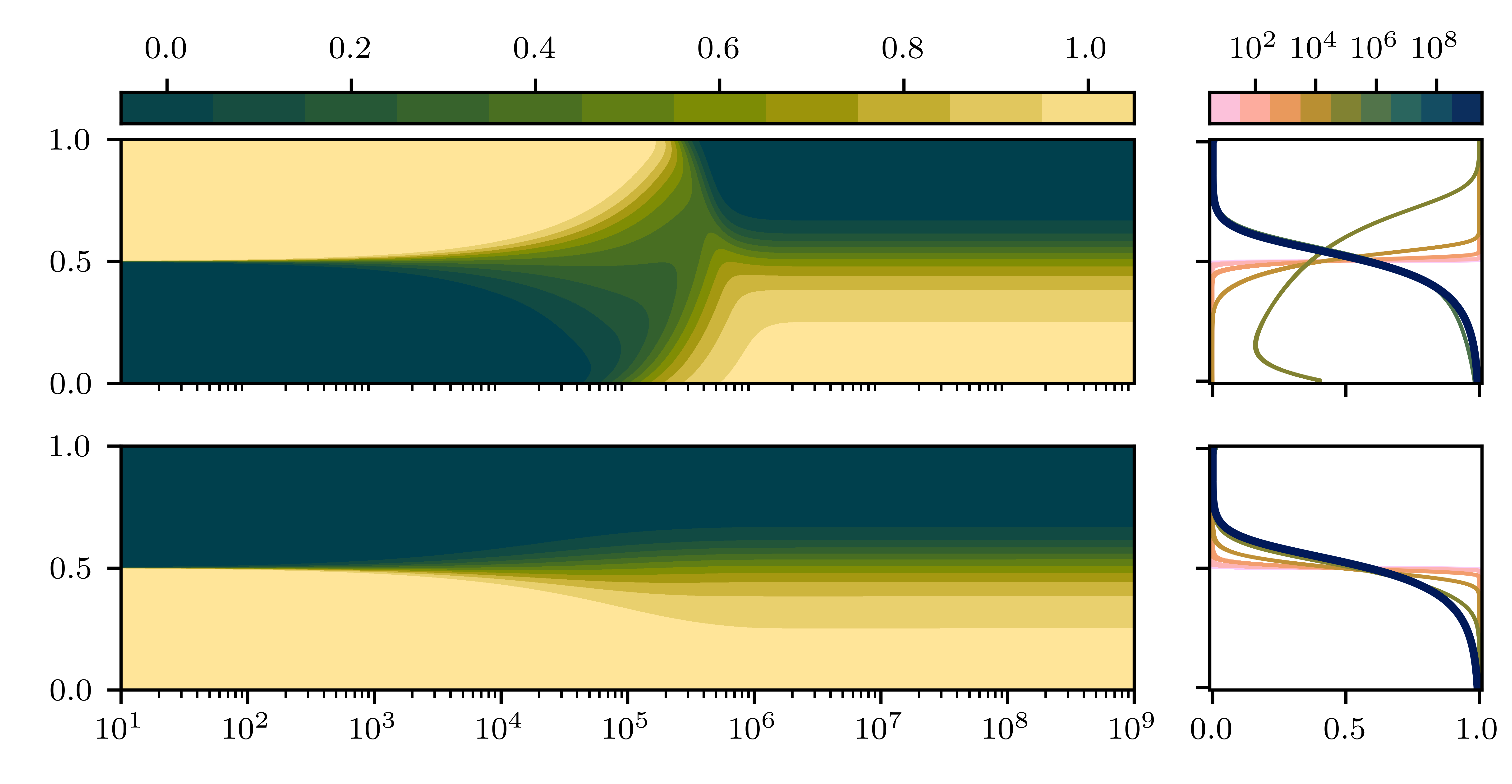}}
\end{center}
\vspace{-0.5cm}
%%% where xxxxxx name represents "figurename.eps"
\caption{Example of numerical solutions for $R_d=2.0$. Evolution of the small-particle concentration field $\phi_{s}(\hat{t},\hat{z})$ with nondimensional height $\hat{z}=z/h$ and nondimensional time $\hat{t}=t\,[h/f_{sf}]^{-1}$, together with the final concentration profiles, for a granular flow that is initially (A) unstably segregated and (B) stably segregated.}
\label{fig:2}
\end{figure}

For the present study, the model was not coupled with a momentum equation, as done previously \cite{Trewhela24,TrewhelaUlloa2024}. This coupling was deemed unnecessary since transient momentum transfer does not control the final mixing state, nor does it govern the segregation–diffusion balance. This was evidenced by analysis of the energetic balance in previous work \cite{TrewhelaUlloa2024}. The solution sets consisted of varying the size ratio $R_d$ for two different initial conditions (stable and unstable) of the small-particle distribution $\phi_s$. A first set of simulations was performed for an inversely graded (unstable) initial condition with $R_d \in [1.1:0.1:3.2]$, and a second set for a normally graded (stable) initial condition with $R_d \in [1.1:0.1:5.0]$, yielding a total of 62 simulations. For this range of $R_d$ and the parameter values listed in table~\ref{tab:params}, the Péclet number spans $0.8 \leq Pe \leq 200$. This range is broad enough to capture the transition from diffusion-dominated remixing to segregation-dominated transport. In all cases, the parameters in table~\ref{tab:params} were held fixed, as was the initial small--large particle interface, which was placed at $z=h/2$, or $\hat{z}=1/2$. As an example, figure~\ref{fig:2} shows the numerical solutions for $R_d=2.0$ for both the unstable and stable initial conditions. These results show two distinctive features of our solutions: (i) the final mixed layer is independent of the initial condition, a feature already observed in the literature \cite{Gray06,Trewhela24}; and (ii) the unstable case shows an intermediate (apparent) mixing state, not present in stable conditions, but it is also a well-known feature of the inversely graded initial condition problem \cite{Gray2018}. Remarkably, this apparently mixed state does not necessarily translate into a wider or narrower mixing-layer, as we will explore later. 

Guided by the theoretical model and the numerical simulations, we next examine three key aspects of the sheared granular flow. First, we quantify the degree of segregation, the reciprocal quantity of the degree of mixing \cite{Danckwerts1952,TrewhelaUlloa2024}. Secondly, we characterize the bidisperse mixing-layer and its dependence on the governing control parameters. Finally, we investigate the relevant timescales of the segregation--remixing dynamics, with particular emphasis on the conditions under which the system attains maximum mixing and a stable, maximally segregated state. Throughout this study we analyze these quantities as functions of the particle-size ratio $R_{d}$ and the segregation P\'eclet number.

%%%%%%%%%%%%%%%%%%%%%
\vspace{0cm}
\section{Degree of segregation}\label{sec:3}

For a granular flow, it is possible to quantify how well segregated or mixed the flow is by introducing the degree of segregation (and mixing). Following \cite{Danckwerts1952,TrewhelaUlloa2024}, the degree of segregation in a bidisperse granular column is quantified as
\begin{equation}
\mathscr{S}_\phi
=
\left(\frac{\sigma_\phi}{\sigma_{sg}}\right)^{2}
=
\frac{1}{\sigma_{sg}^{2}}
\left[
\mathbb{E}\left(\phi_s^{2}\right)
-
\mathbb{E}\left(\phi_s\right)^{2}
\right],
\end{equation}
where $\phi_s(\hat z,\hat t)$ is the local volume fraction of small grains, $\sigma_\phi^2$ is its variance, and $\sigma_{sg}^{2}=1/4$ is the reference variance for a fully segregated state. Here, $\hat{z}=z/h$ is the nondimensional height and $\hat{t}=t [h/f_{sf}]^{-1}$ is the nondimensional time, scaled by the segregation-flux timescale. This scaling assumes that segregation makes an $O(1)$ contribution to the evolution of the degree of segregation. \textcolor{black}{The operator $\mathbb{E}$ denotes the expected value and, in this context, corresponds to a depth average over the nondimensional column:
\begin{equation}
\mathbb{E}[f]
=
\int_0^1 f(\hat z,\hat t)\,{\rm d}\hat z,
\qquad \hat z\in[0,1].
\end{equation}
Thus, $\mathscr{S}_{\phi}(t)$ is a global quantity that only depends in time.}

Assuming sufficient regularity of the fields, the temporal evolution of $\mathscr{S}_\phi$ is governed by the second moment,
\begin{equation}\label{eq:seg}
\frac{{\rm d} \mathscr{S}_\phi}{{\rm d} \hat t}
=
\frac{1}{\sigma_{sg}^{2}}
\frac{{\rm d}}{{\rm d}\hat t}\mathbb{E}\left(\phi_s^{2}\right)
=
\frac{1}{\sigma_{sg}^{2}}
\mathbb{E}\left(\frac{\partial \phi_s^{2}}{\partial \hat t}\right).
\end{equation}

In turn, the evolution of $\phi_s$ is governed by the nondimensional segregation–diffusion equation
\begin{equation}\label{eq:phi_s_ndl}
\frac{\partial \phi_s}{\partial \hat t}
=
\frac{\partial}{\partial \hat z}
\left[
\hat f_{sl}\,\phi_s(1-\phi_s)
+
\frac{\hat f_{sl}}{Pe}\frac{\partial \phi_s}{\partial \hat z}
\right],
\end{equation}
where $\hat{f}_{sl}(\hat{z},\hat{t})$ is the nondimensional segregation flux and $Pe$ is a Péclet number. Multiplying \eqref{eq:phi_s_ndl} by $2\phi_s$, we obtain the evolution equation for the quadratic field $\phi_s^2$:
\begin{equation}\label{eq:phi2_s}
\frac{\partial \phi_s^2}{\partial \hat t}
=
\frac{\partial}{\partial \hat z}
\left[
2\hat f_{sl}\phi_s^2(1-\phi_s)
+
\frac{\hat f_{sl}}{Pe}\frac{\partial \phi_s^2}{\partial \hat z}
\right]
-\hat f_{sl}(1-\phi_s)\frac{\partial \phi_s^2}{\partial \hat z}
-\frac{2\hat f_{sl}}{Pe}\left(\frac{\partial \phi_s}{\partial \hat z}\right)^2.
\end{equation}

Substituting \eqref{eq:phi2_s} into \eqref{eq:seg} and integrating over the domain yields
\begin{equation}\label{eq:evolution-eq-seg}
\frac{{\rm d}}{{\rm d}\hat t}\mathbb{E}(\phi_s^2)
=
\left[
2\hat f_{sl}\phi_s^2(1-\phi_s)
+
\frac{\hat f_{sl}}{Pe}\frac{\partial \phi_s^2}{\partial \hat z}
\right]_{\hat z=0}^{\hat z=1}
-\int_0^1 \hat f_{sl}(1-\phi_s)\frac{\partial \phi_s^2}{\partial \hat z}\,{\rm d}\hat z
-\int_0^1 \frac{2\hat f_{sl}}{Pe}\left(\frac{\partial \phi_s}{\partial \hat z}\right)^2 {\rm d}\hat z.
\end{equation}
The first term represents boundary flux contributions, while the integral terms capture bulk production and dissipation of variance. In particular, the last term is strictly positive definite and represents the shear-controlled diffusive smoothing of particle concentration gradients -- or mixing rate. For no-flux boundary conditions, the degree of segregation is governed by the following evolution equation:
\begin{equation}\label{eq:evolution-eq-seg-2}
\frac{{\rm d} \mathscr{S}_\phi}{{\rm d} \hat t}
=
-\frac{1}{\sigma_{sg}^{2}}
\int_0^1 \hat f_{sl}(1-\phi_s)\frac{\partial \phi_s^2}{\partial \hat z}\,{\rm d}\hat z
-\frac{1}{\sigma_{sg}^{2}}
\int_0^1 \frac{2\hat f_{sl}}{Pe}\left(\frac{\partial \phi_s}{\partial \hat z}\right)^2 {\rm d}\hat z.
\end{equation}

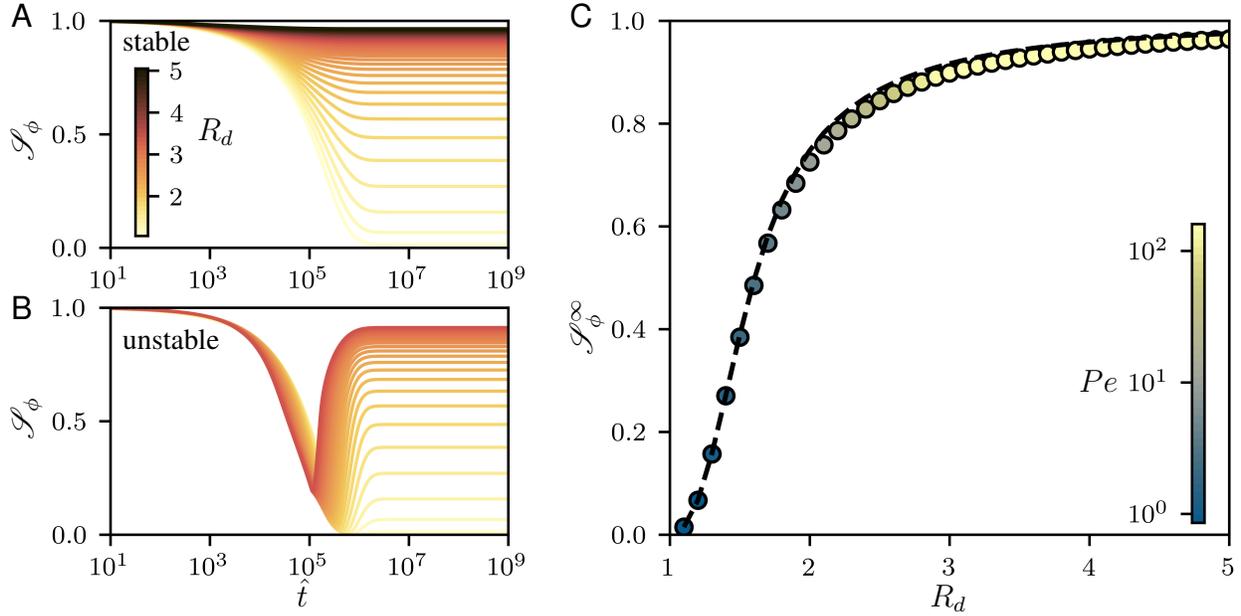
\begin{figure}[!h]
\begin{center}
\SetLabels
\L (0.0*0.97) \sf{A}\\
\L (0.0*0.48) \sf{B}\\
\L (0.45*0.97) \sf{C}\\
\L (0.86*0.36) $Pe$\\
\L (0.0*0.745) \rotatebox{90}{$\mathscr{S}_{\phi}$}\\
\L (0.0*0.28) \rotatebox{90}{$\mathscr{S}_{\phi}$}\\
\L (0.23*0.0) $\hat{t}$\\
\L (0.74*0.0) $R_{d}$\\
\L (0.09*0.93) \textcolor{black}{\sf stable}\\
\L (0.09*0.45) \textcolor{black}{\sf unstable}\\
\L (0.15*0.78) \textcolor{black}{$R_d$}\\
\L (0.45*0.42) \rotatebox{90}{$\mathscr{S}^{\infty}_{\phi}$}\\
\endSetLabels
\strut\AffixLabels{\includegraphics[width=\textwidth]{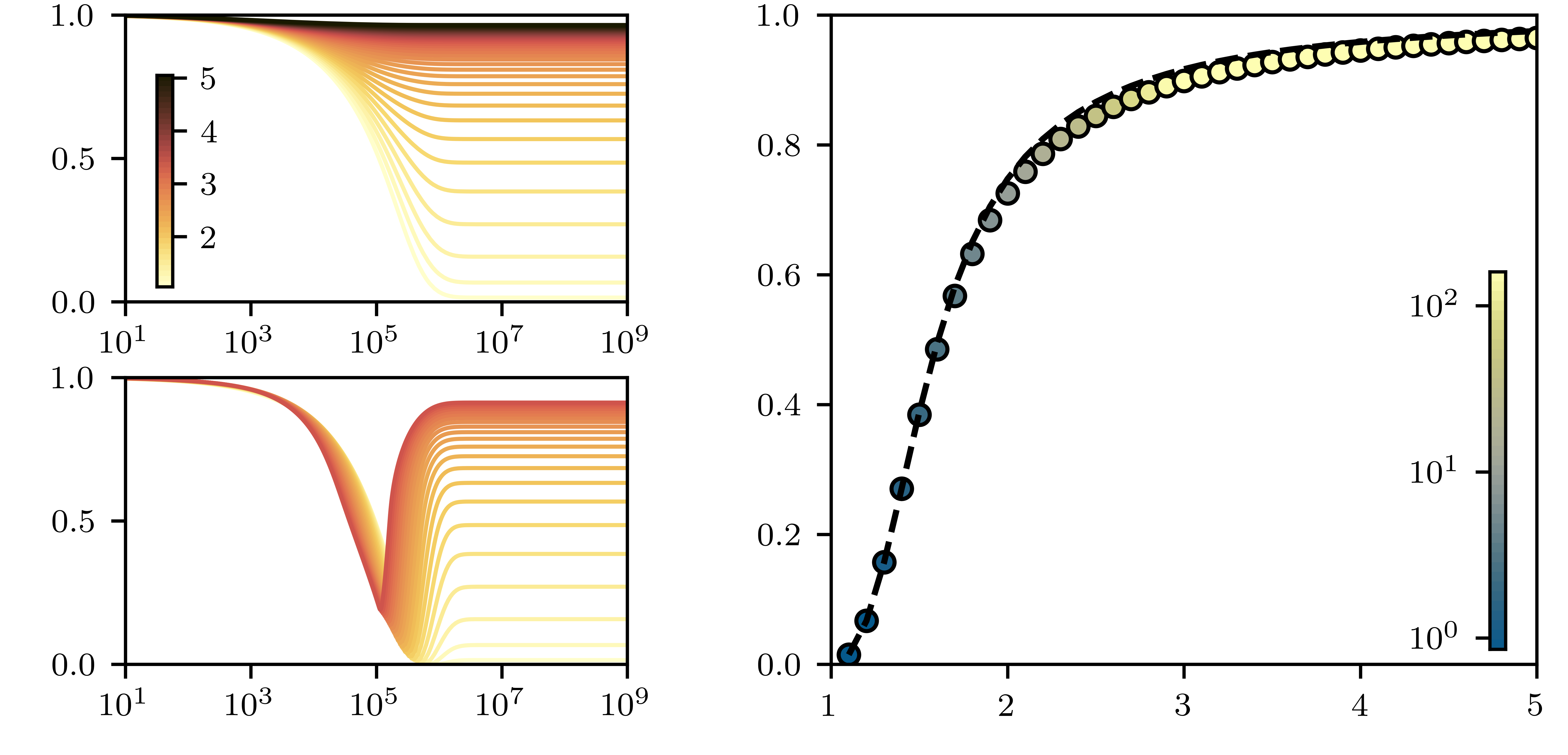}}
\end{center}
\vspace{-0.5cm}
\caption{Degree of segregation $\mathscr{S}_\phi$ as a function of particle-size ratio $R_d$ and segregation Péclet number $Pe$. Temporal evolution of $\mathscr{S}_\phi$ for $1 \leq R_d \leq 5$, starting from an (A) initially stable segregated state and an (B) initially unstable segregated state. (C) Final state of the degree of segregation, denoted as $\mathscr{S}^{\infty}_\phi$ as a function of $R_d$ and $Pe$, with colour indicating the magnitude of $Pe$. The dashed line traces the analytical solution for the asymptotic, steady-state degree of segregation in equation~\eqref{eq:SPe} with $Pe$ evaluated according to equation~\eqref{eq:Peclet}.}
\label{fig:3}
\end{figure}

From the numerical integration of the convection--diffusion equation \eqref{eq:evolution-eq-seg}, we resolve the temporal evolution of the degree of segregation, $\mathscr{S}_\phi$. Figure~\ref{fig:3}{\sf A--B} shows the evolution of $\mathscr{S}_\phi$ for systems that are initially stably segregated, or normally graded, with large particles above small particles, and initially unstably segregated, or inversely graded, with small particles above large particles. In the former case, starting from a fully segregated, normally graded state leads to a decrease in $\mathscr{S}_\phi$, which asymptotically approaches a minimally segregated state. By contrast, figure~\ref{fig:3}{\sf A} shows that, when a bidisperse granular flow starts from an unstably segregated condition, $\mathscr{S}_\phi$ reaches a minimum before converging to a new stable, but partially segregated, equilibrium state. We find that, in all cases and independently of the initial condition, the steady-state degree of segregation $\mathscr{S}^{\infty}_{\phi}$ increases with increasing $R_d$, as expected and as shown in figure~\ref{fig:3}{\sf C}. However, $\mathscr{S}_{\phi}$ does not reach unity, even for relatively large size ratios, consistent with the asymptotic steady-state analytical solution, shown by the dashed line in figure~\ref{fig:3}{\sf C}, given by
\begin{equation}
\mathscr{S}_{\phi}=1-\frac{4}{Pe}\,\tanh\left[\frac{Pe}{4}\right].
\label{eq:SPe}
\end{equation}
Beyond $R_d=5$, other competing phenomena, such as spontaneous percolation \citep{Wilkinson1982} or reverse segregation \citep{Thomas18}, may become relevant; thus, complete segregation in the presence of diffusion is unlikely in real systems, as also shown in experiments and computational simulations \citep{ferdowsi17,Duan2024,zhao2025scaling}.

% ------ Result sections [not calling them results] ----------

\section{Bidisperse mixing-layer}
\label{sec:4}

We seek a measure of the vertical extent of the bidisperse mixing region, namely the interval over which small and large grains coexist in appreciable relative proportions. To analyse the mixing region at steady state, we start from the self-similar concentration profile of small particles at steady state \cite{Gray06,Trewhela24,TrewhelaUlloa2024},
\begin{equation}
    \phi_s(\hat z)
    =
    \frac{e^{Pe(\mathcal{K}-\hat z)}}{1+e^{Pe(\mathcal{K}-\hat z)}},
    \label{eq:phisan}
\end{equation}
where $\mathcal{K}=1/2$ denotes the depth-averaged concentration of small particles and coincides with the position of the small and large particles interface. It is reflected in equation~\eqref{eq:phisan} that, physically, large $Pe$ implies segregation-dominated transport and therefore a sharp transition between small- and large-particle-rich regions, whereas small $Pe$ implies stronger diffusive smoothing and hence a broader coexistence region.

A natural definition of the mixing-layer thickness is obtained by centering the layer at the location of maximum concentration gradient and taking its boundaries as the two points of strongest curvature on either side. For the profile \eqref{eq:phisan}, the first derivative is
\begin{equation}
    \frac{\partial \phi_s}{\partial \hat z}
    =
    -\,Pe\frac{e^{Pe(\mathcal{K}-\hat z)}}
    {\left[1+e^{Pe(\mathcal{K}-\hat z)}\right]^2}
    =
    -\,Pe\,\phi_s(1-\phi_s).
    \label{eq:dphidz_mix}
\end{equation}
Hence $|\partial \phi_s/\partial \hat z|$ is maximal when $\phi_s=1/2$, which occurs at $\hat z=\mathcal{K}$. The mixing-layer is therefore centred at $\hat z=\mathcal{K}$. To quantify its width, we define the layer boundaries as the two symmetric locations at which $\partial_{\hat z\hat z}\phi_s$ attains an extremum, that is, where the profile bends most strongly about the point of maximum slope. For constant $Pe>0$, the third derivative is
\begin{equation}
    \frac{\partial^3 \phi_s}{\partial \hat z^3}
    =
    -\,\frac{Pe^3\,e^{Pe(\mathcal{K}-\hat z)}
    \left[1-4e^{Pe(\mathcal{K}-\hat z)}+e^{2Pe(\mathcal{K}-\hat z)}\right]}
    {\left[1+e^{Pe(\mathcal{K}-\hat z)}\right]^4}.
    \label{eq:d3phidz3_mix}
\end{equation}
The extrema of $\partial_{\hat z\hat z}\phi_s$ therefore satisfy $$1-4e^{Pe(\mathcal{K}-\hat z)}+e^{2Pe(\mathcal{K}-\hat z)}=0.$$ Introducing $\zeta=e^{Pe(\mathcal{K}-\hat z)}$, the latter equation reduces to a quadratic equation `$\zeta^2-4\zeta+1=0$' whose roots are $\zeta=2\pm\sqrt{3}$. The corresponding vertical locations are
\begin{equation}
    \hat z_1=\mathcal{K}-\frac{1}{Pe}\ln(2+\sqrt{3}),
    \qquad
    \hat z_2=\mathcal{K}+\frac{1}{Pe}\ln(2+\sqrt{3}),
    \label{eq:z1z2_mix}
\end{equation}
which are symmetric about $\hat z=\mathcal{K}$, consistent with the symmetry of the self-similar profile. We therefore define the characteristic bidisperse mixing-layer thickness as
\begin{equation}
    \hat{\delta}_m
    =
    \hat z_2-\hat z_1
    =
    \frac{2}{Pe}\ln(2+\sqrt{3}).
    \label{eq:delta_m}
\end{equation}
Thus the thickness scales inversely with the P\'eclet number,
\begin{equation}\label{eq:delta_Pe_ctn}
    \hat{\delta}_m \approx \frac{2.634}{Pe}.
\end{equation}
This inverse scaling makes the underlying physics transparent. Increasing $Pe$ strengthens segregation relative to remixing, sharpens the concentration interface, and compresses the coexistence region. In contrast, decreasing $Pe$ has the opposite effect and broadens the mixing-layer. The limiting case occurs when the predicted mixing-layer thickness spans the entire domain, $\hat{\delta}_m=1$. This condition defines a critical Péclet number, $Pe_c\approx2.634$. Using the closure for the Péclet number, the corresponding critical particle-size ratio $R_c$ satisfies 
\begin{equation}
    Pe_{c}\approx 2.634\approx \frac{\mathcal{B}\rho_{*}gh}{\mathcal{A}\,p_0}(R_c-1)\left[1+\frac{1}{2}\mathcal{E}(R_c-1)\right].
    \label{eq:Rc}
\end{equation}
Solving this quadratic equation gives $R_c\approx1.27$, together with a negative root, $R_c\approx-0.22$, which has no physical meaning because $R_d>1$ by definition. Thus, $R_c$ provides an approximate threshold: for size ratios below this value, remixing can produce a domain-spanning mixing-layer, whereas for larger size ratios, segregation is expected to maintain a sharper interface.

This estimate is relevant for DEM simulations of dense granular flows. Slight particle-size polydispersity is often introduced to suppress crystallization, layering, and other numerical artifacts \citep{Guillard2016,Duan2024,Singh24,zhao2025scaling}. However, the estimate $R_c\approx1.27$ suggests that even modest size contrast may influence the macroscopic mixing--segregation dynamics. In particular, values near $R_d\approx1.2$ may already produce measurable segregation in some parameter regimes. Therefore, DEM studies intended to approximate monodisperse flows should verify that the imposed polydispersity does not introduce unintended size-segregation effects.

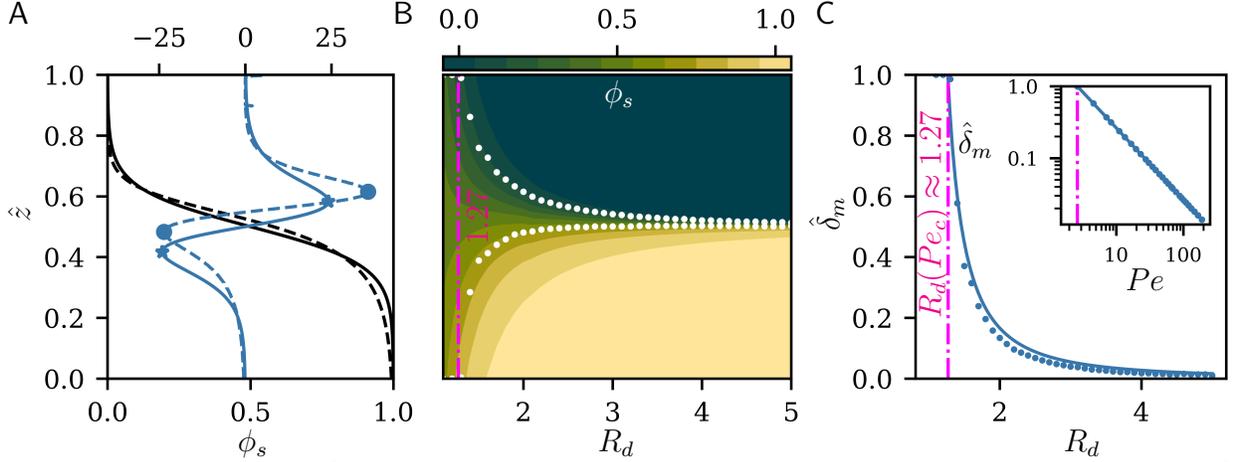
\begin{figure}[!h]
\begin{center}
\SetLabels
\L (0.0*0.99) $\sf A$\\
\L (0.31*0.99) $\sf B$\\
\L (0.65*0.99)  $\sf C$\\
\L (0.0*0.5) \rotatebox{90}{$\hat{z}$}\\
\L (0.645*0.48) \rotatebox{90}{\textcolor{black}{$\hat{\delta}_m$}}\\
%\L (0.765*0.65) \textcolor{black}{$\hat{\delta}_m$}\\
\L (0.90*0.35) \textcolor{black}{$Pe$}\\
\L(0.725*0.98){\scriptsize\textcolor{black}{$-\cdot \,\, R_{d}(Pe_c)\approx1.27$}}\\
\L(0.725*0.91){\color{NavyBlue}{$\cdot\cdot\cdot \,$ \scriptsize numerical result}}\\
\L(0.725*0.86){\color{NavyBlue}{\rule{0.4cm}{0.8pt}}}\\
\L(0.765*0.85){\scriptsize\color{NavyBlue}{ Eq.~\eqref{eq:delta_Pe_ctn}};}\\
\L(0.85*0.86){\color{black}{\rule{0.4cm}{0.8pt}}}\\
\L(0.88*0.85){\scriptsize\color{black}{ Eq.~\eqref{eq:delta_m_Pe_eff}}}\\
\L (0.37*0.45) \rotatebox{90}{\textcolor{black}{$1.27$}}\\
% \L (0.68*0.647) \textcolor{white}{$\Bigg\updownarrow$}\\
% \L (0.68*0.255) \textcolor{white}{$\Bigg\updownarrow$}\\
\L (0.16*0.96) $ \partial_{\hat{z}\hat{z}}\phi_{s}$\\
\L (0.185*0.0) $\phi_{s}$\\
\L (0.475*0.0) $R_d$\\
\L (0.48*1.0) $\textcolor{black}{\phi_s}$\\
\L (0.85*0.0) $R_d$\\
\endSetLabels
\strut\AffixLabels{\includegraphics[width=\textwidth]{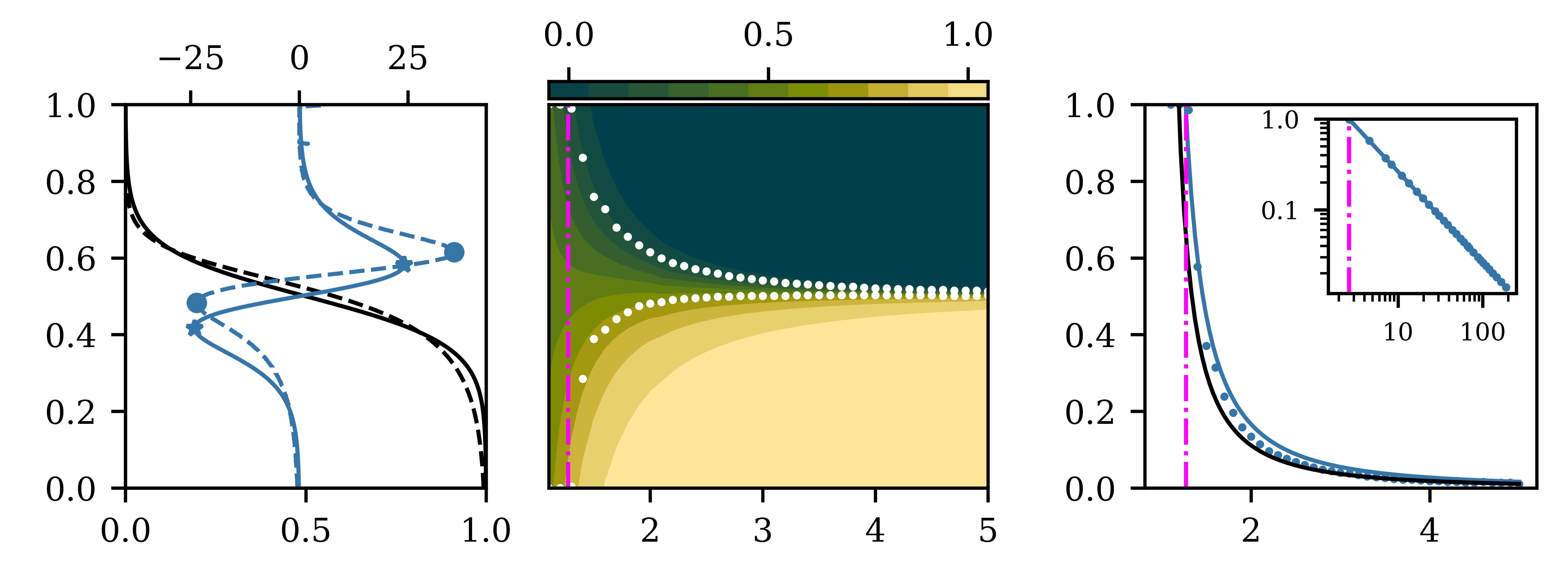}}
\end{center}
\vspace{-0.5cm}
\caption{Mixed-layer thickness $\hat{d}_m$ as a function of $R_d$ and $Pe$. (A) Illustration of the approach used to calculate $\hat{\delta}_m$ for $R_d=2$, i.e. $Pe=15.85$. The black solid line corresponds to the equilibrium small-particle concentration profile for $\phi_s$ in equation~\eqref{eq:phisan}, and the blue line to its second derivative, $\partial_{\hat{z}\hat{z}}\phi_s$. Circular markers denote the heights where $\partial_{zzz}\phi_s=0$; the distance between these heights defines the bidisperse mixing-layer thickness, $\hat{\delta}_m$. (B) Final profiles of $\phi_s$ for various $R_d$ values, with white dots marking the positions where $\partial_{\hat{z}\hat{z}\hat{z}}\phi_s=0$. (C) $\hat{\delta}_m$ as a function of $R_d$ and $Pe$ (inset); solid lines show the theoretical prediction equation~\eqref{eq:delta_Pe_ctn}.}
\label{fig:4}
\end{figure}

Figure~\ref{fig:4} shows how we define the mixing-layer thickness and how it varies with the particle-size ratio $R_d$ and the Péclet number $Pe$. 
The steady-state theoretical solution for the small-particle concentration, $\phi_s$ in equation~\eqref{eq:phisan}, is compared with the numerical solution for $R_d=2.0$ in figure~\ref{fig:4}{\sf A}. Consistent with figure~\ref{fig:3}, this steady profile is independent of the initial condition. To identify the bounds of the mixing-layer, we plot the second derivative, $\partial_{\hat{z}\hat{z}}\phi_s$, in blue. Changes in curvature locate the upper and lower edges of the transition region, marked by stars $\star$ for the theoretical solution and circles $\bullet$ for the numerical solution. Although the theoretical and numerical concentration profiles show small differences, which propagate into $\partial_{\hat{z}\hat{z}}\phi_s$ and the precise locations of $\hat{z}_1$ and $\hat{z}_2$, the resulting thickness $\hat{\delta}_m=\hat{z}_2-\hat{z}_1$ differs by less than 5\% on average across our data. Thus, the layer thickness is robust even when the absolute position of the transition region shifts slightly within the sheared layer.

Using this curvature-based method, we compute $\hat{\delta}_m$ for all numerical steady-state profiles shown in figure~\ref{fig:4}{\sf B}. The white circles mark the identified bounds, $\hat{z}_1$ and $\hat{z}_2$, over the filled contours of $\phi_s$. As $R_d$ increases, these bounds converge, indicating that $\hat{\delta}_m\rightarrow0$ as segregation strengthens and the interface sharpens. Conversely, for small size ratios, the mixing-layer spans the entire domain, $\hat{\delta}_m=1$, indicating saturated mixing. The critical size ratio associated with this transition, $R_c(Pe_c)\approx1.27$, is obtained from the quadratic relation in equation~\eqref{eq:Rc} and is shown by the vertical dash-dotted line in figure~\ref{fig:4}{\sf B}. 
We compare the numerical values of $\hat{\delta}_m$ with the theoretical prediction in equation~\eqref{eq:delta_Pe_ctn} as a function of $R_d$ in figure~\ref{fig:4}{\sf C}; the inset shows the corresponding segregation Péclet number, $Pe=(\mathcal{B}\rho_*\,g\,h)/(\mathcal{A}\,p_0)[(R_d-1)+\frac{1}{2}\mathcal{E}(R_d-1)^2].$ Despite small discrepancies in the individual estimates of $\hat{z}_1$ and $\hat{z}_2$, the theory captures both the trend and magnitude of the numerical mixing-layer thickness, \textcolor{black}{especially for small and large values of $R_{d}$}. 

This behavior is physically consistent with the competition between segregation and diffusive remixing. For smaller values of $R_d$, size-selective segregation is weak, the degree of segregation remains low, and remixing can spread the coexistence region across the full layer height, leading to $\hat{\delta}_m\rightarrow1$. For larger values of $R_d$, segregation becomes more efficient, the concentration interface sharpens, and the mixing-layer narrows as $\hat{z}_1\rightarrow\hat{z}_2$. The critical value $Pe_c\approx2.634$ therefore provides a useful threshold for predicting when the mixing-layer becomes comparable to the full height of the sheared granular layer.

To provide a more general description of the mixing-layer, we also consider an alternative steady-state solution for the small-particle concentration $\phi_s$ that accounts for lithostatic pressure. \textcolor{black}{In this case, the pressure increases linearly with depth, or equivalently decreases linearly with height, so that $p\propto(1-\hat z)$. We therefore consider a closure in which the local segregation strength depends on both depth and concentration:}
\begin{equation}
Pe(\hat{z})
=
\frac{\mathcal{B}(R_d-1)}{\mathcal{A}\Phi(1-\hat{z})}
\left[1+\mathcal{E}(R_d-1)(1-\phi_s)\right],
\label{eq:Pe(z)}
\end{equation}
where $\mathcal{A}$, $\mathcal{B}$, and $\Phi$ are constants (table~\ref{tab:params}), $\mathcal{E}$ is a dimensionless coupling parameter, and $R_d>1$ is the particle-size ratio. This formulation captures two effects: segregation is enhanced by increasing particle-size contrast, and its local strength varies with the concentration field through the depletion of small particles. As a result, the segregation dynamics become nonlinear, and the steady concentration profile cannot be obtained in closed form.

To obtain a local estimate of the layer thickness, we approximate the concentration profile near the centre of the layer by a logistic form with a locally defined effective Péclet number,
\begin{equation}
    \phi_s(\hat z)
    \approx
    \frac{\exp\!\left[q(\hat z)\right]}
    {1+\exp\!\left[q(\hat z)\right]},
    \qquad
    q(\hat z)=Pe(\hat z)\,(\mathcal{K}-\hat z).
    \label{eq:phi_feedback2}
\end{equation}
Here $q(\hat z)$ should be interpreted as a local approximation, not as an exact self-similar solution. Differentiating gives
\begin{equation}
    \frac{\partial \phi_s}{\partial \hat z}
    =
    q'(\hat z)\,\phi_s(1-\phi_s).
\end{equation}
Provided $q'(\hat z)$ varies slowly across the transition region, the maximum gradient still occurs near $\phi_s=1/2$, so that the layer centre remains close to $\hat z_c\approx \mathcal{K}$. Evaluating \eqref{eq:Pe(z)} at $\hat z=\mathcal{K}$ and $\phi_s=1/2$ yields the effective segregation strength
\begin{equation}
    Pe_{\mathrm{eff}}
    =
    \frac{\mathcal{B}(R_d-1)}{\mathcal{A}\Phi(1-\mathcal{K})}
    \left[1+\frac{1}{2}\mathcal{E}(R_d-1)\right].
    \label{eq:Pe_eff_feedback2}
\end{equation}
Under this local approximation, the mixing-layer thickness follows from the constant-\(Pe\) result,
\begin{equation}
    \hat{\delta}_m
    \approx
    \frac{2}{Pe_{\mathrm{eff}}}\ln(2+\sqrt{3}),
\end{equation}
so that
\begin{equation}\label{eq:delta_m_Pe_eff}
    \hat{\delta}_m
    \approx
    \frac{2\mathcal{A}\Phi(1-\mathcal{K})\,\ln(2+\sqrt{3})}
    {\mathcal{B}(R_d-1)\left[1+\tfrac{1}{2}\mathcal{E}(R_d-1)\right]}.
    %\label{eq:delta_feedback2}
\end{equation}

Equation \eqref{eq:delta_m_Pe_eff} shows that the mixing-layer thickness decreases as the size ratio increases, but now with an additional nonlinear correction arising from concentration-dependent feedback. In the limit $\mathcal{E}\to 0$, the classical scaling $\hat{\delta}_m\sim (R-1)^{-1}$ is recovered. For finite \(\mathcal{E}\), however, the feedback enhances segregation in regions where small particles are locally depleted, which further sharpens the concentration gradients and thins the coexistence region. In the large-$R_d$ limit, this gives the stronger asymptotic scaling $\hat{\delta}_m\sim (R-1)^{-2}$.

\textcolor{black}{The prediction for the mixing-layer thickness obtained with depth-dependent pressure is shown by the black solid line in figure~\ref{fig:4}C. This correction is present but small. As expected, the estimate based on the effective Péclet number in equation~\eqref{eq:delta_m_Pe_eff} converges to the constant-pressure prediction in equation~\eqref{eq:delta_Pe_ctn} at large $R_d$. Physically, this occurs because stronger segregation sharpens the concentration interface, so that $\hat{z}_1\rightarrow\hat{z}_2$ and the pressure variation across the mixing-layer becomes negligible. Larger differences appear near the critical values of $Pe$ and $R_d$, where the mixing-layer is broad and the pressure contrast between its upper and lower bounds is largest. Overall, however, the depth-dependent scaling remains close to the numerical results, supporting the robustness of the constant-pressure approximation while quantifying the correction introduced by lithostatic pressure.}

\section{Mixing--Segregation timescales}\label{sec:5}

Finally, we analyse the characteristic timescales governing  mixing and segregation states in the granular flow, with particular attention to the approach to two limiting states: the maximally mixed state for granular systems that are initially stably segregated and systems that are initially unstably segregated. We combine the time-dependent numerical solutions with the asymptotic steady-state predictions to quantify how rapidly the system approaches each limit.

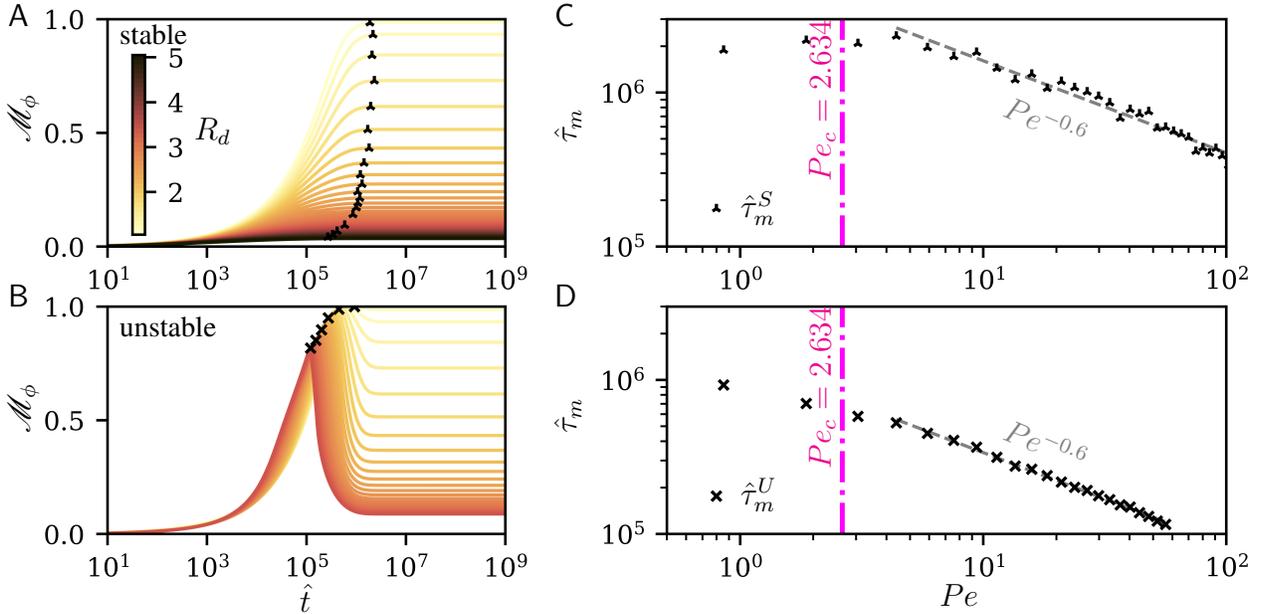
\begin{figure}[!h]
\begin{center}
\SetLabels
\L (0.0*0.97) $\sf A$\\
\L (0.0*0.5) $\sf B$\\
\L (0.44*0.97)  $\sf C$\\
\L (0.44*0.5)  $\sf D$\\
\L (0.0*0.775) \rotatebox{90}{$\mathscr{M}_{\phi}$}\\
\L (0.0*0.29) \rotatebox{90}{$\mathscr{M}_{\phi}$}\\
\L (0.44*0.775) \rotatebox{90}{\textcolor{black}{$\hat{\tau}_{m}$}}\\
\L (0.44*0.29) \rotatebox{90}{\textcolor{black}{$\hat{\tau}_{m}$}}\\
\L (0.645*0.71) \rotatebox{90}{\textcolor{black}{$Pe_c=2.634$}}\\
\L (0.645*0.23) \rotatebox{90}{\textcolor{black}{$Pe_c=2.634$}}\\
\L (0.8*0.82) \rotatebox{-23}{\textcolor{gray}{$Pe^{-0.6}$}}\\
\L (0.8*0.28) \rotatebox{-23}{\textcolor{gray}{$Pe^{-0.6}$}}\\
\L (0.59*0.65) \textcolor{black}{$\hat{\tau}^{S}_{m}$}\\
\L (0.59*0.17) \textcolor{black}{$\hat{\tau}^{U}_{m}$}\\
\L (0.235*0.0) $\hat{t}$\\
\L (0.75*0.0) $Pe$\\
\L (0.09*0.94) \textcolor{black}{\sf stable}\\
\L (0.09*0.46) \textcolor{black}{\sf unstable}\\
\L (0.15*0.78) \textcolor{black}{$R_d$}\\
\endSetLabels
\strut\AffixLabels{\includegraphics[width=\textwidth]{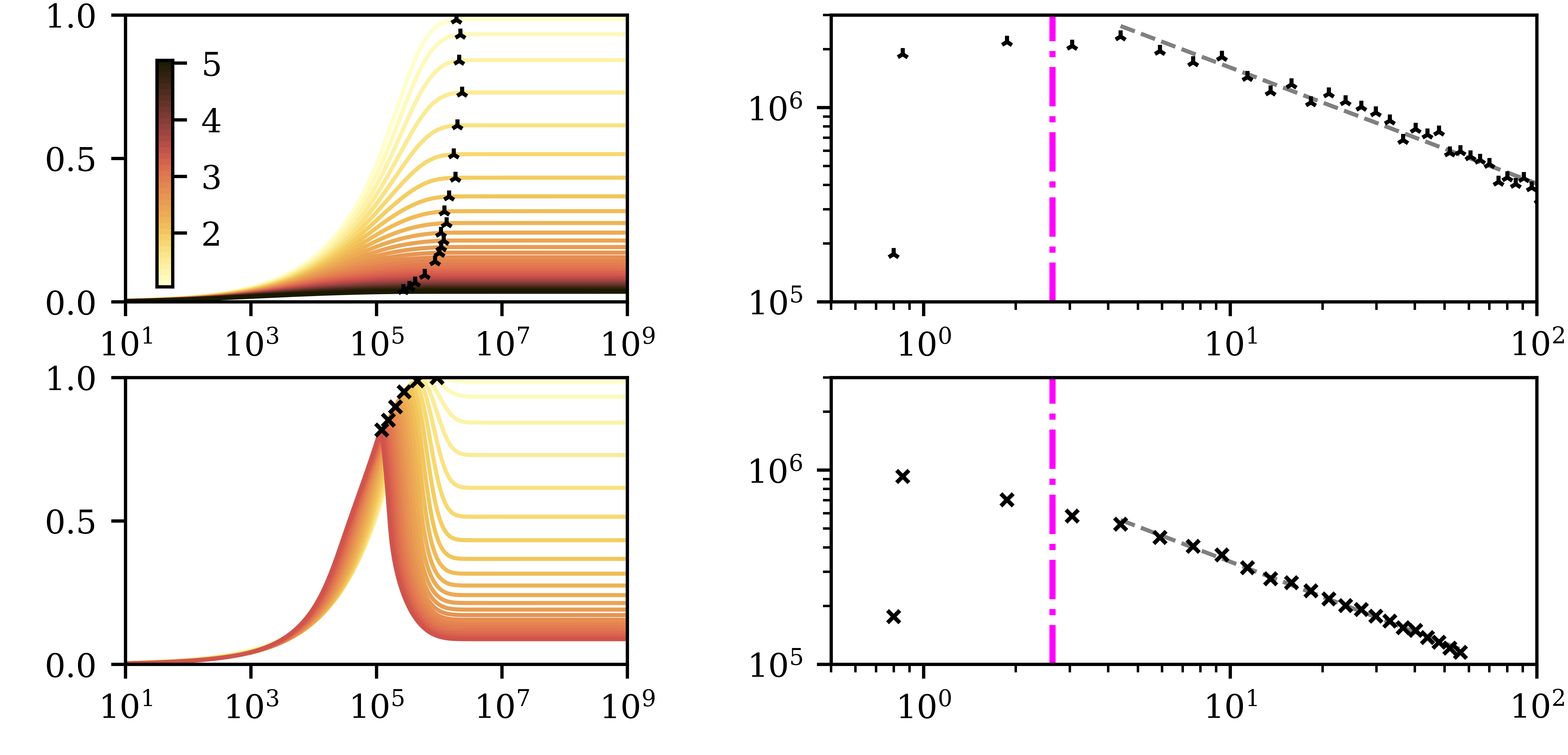}}
\end{center}
\vspace{-0.5cm}
\caption{Timescale for maximal mixing. Temporal evolution of $\mathscr{M}_{\phi}$ for $1 \leq R_d \leq 5$, starting from (A) an initially stable segregated state and (B) an initially unstable segregated state. Markers denote the time $\hat{\tau}_m$ at which the maximum degree of mixing, $\mathscr{M}_{\phi}=1-\mathscr{S}_{\phi}$, is achieved, computed from the minimum value of $\mathscr{S}_{\phi}(t)$. $\hat{\tau}_m$ as a function of $Pe$ for (C) an initially stable segregated state and (D) an initially unstable segregated state.}
\label{fig:5}
\end{figure}

For each case, we track the degree of mixing, $\mathscr{M}_\phi=1-\mathscr{S}_\phi$, which provides a bulk measure of concentration variance across the layer: high values of $\mathscr{M}_\phi$ correspond to enhanced mixing, whereas low values indicate stronger segregation. We first determine how the control parameters $R_d$ and $Pe$ influence the time $\hat{\tau}_{m}$ required for $\mathscr{M}_\phi$ to attain its maximum value, corresponding to the state of maximum mixing. Results are shown in figure~\ref{fig:5}A-B. From these results, it is straightforward to estimate the time at which maximal mixing is attained as a function of $R_d$ for the initially unstable granular arrangement, as can be shown by the cross markers in figure~\ref{fig:5}A. For the scenario when the granular system is initially stably segregated, we use an asymptotic solution based on equation~\eqref{eq:evolution-eq-seg-2} to provide an estimation of the time $\hat{\tau}_{m}$ at which the system reaches \(99\%\) of the long-term concentration distribution, denoted by $\mathscr{M}_\phi^{\infty}$. These two timescales characterize the transient competition between diffusive remixing and segregation-driven fluxes and therefore provide a compact description of the temporal response of the system over the $R_{d}$ parameter space. It is apparent that the maximum degree of mixing $\mathscr{M}_{\phi}$ is achieved for smaller values of the particle-size ratio $R_{d}$, which provides a strong control on the segregation flux intensity.

The remaining question is how the characteristic time to maximum mixing, $\hat{\tau}_m$, depends on the P\'eclet number. To address this, we refer to the dimensional convective--segregation--diffusive equation \eqref{eq:phi_nu} and adopt the characteristic scales $h$ for height, $f_{sl}$ for the segregation flux, and $\tau$ for time, such that $z=h\,\hat z$, $\bm{F}_{\Phi_\nu}=f_{sl}\,\hat{\bm F}_{\Phi_\nu}$, and $t=\tau\,\hat t$. For an initially stably segregated configuration, diffusive remixing promotes particle-size homogenisation, whereas segregation opposes this redistribution and tends to restore stratification. Taking $\mathcal{D}_{sl}$ to be height-uniform for scaling purposes, the natural timescale is the diffusive time $\tau=h^{2}/\mathcal{D}_{sl}$. The nondimensional transport equation can then be written as
\begin{equation}\label{eq:nl_phi_nu-2}
\frac{\partial \Phi_\nu}{\partial \hat t}
=
-\frac{\partial}{\partial \hat z}\left[
\underbrace{Pe\,\hat{\bm F}_{\Phi_\nu}\cdot\hat{\bm z}}_{\text{segregation flux}}
-
\underbrace{\frac{\partial \Phi_\nu}{\partial \hat z}}_{\text{diffusive flux}}
\right].
\end{equation}
Equation \eqref{eq:nl_phi_nu-2} makes explicit the competition between segregation and diffusion. Over a unit height, the segregation timescale scales as $\tau_{\mathrm{seg}}\sim Pe^{-1}$, whereas diffusion acts on an $O(1)$ timescale under the present nondimensionalisation. The time $\hat{\tau}_m$ required to attain the maximum-mixing state must therefore lie between the asymptotic limits
\begin{equation}
    \hat{\tau}_m \sim
    \begin{cases}
    \mathrm{const.}, & Pe\ll 1,\\[3pt]
    Pe^{-1}, & Pe\gg 1.
    \end{cases}
    \label{eq:tau_m_scaling}
\end{equation}

Accordingly, any fitted relation of the form $\hat{\tau}_m\propto Pe^{a}$ should be interpreted as an effective crossover scaling, with its exponent restricted to $-1<a<0$. This range reflects the transition from diffusion-dominated remixing at small $Pe$ to segregation-dominated dynamics at large $Pe$. Consistent with this scaling argument, the numerical results in figure~\ref{fig:5}C for initially stably segregated flows show that $\hat{\tau}_m$ is nearly independent of $Pe$ at low $Pe$, but decreases approximately as $\hat{\tau}_m\propto Pe^{-0.6}$ for higher supercritical Péclet numbers over almost two decades.

For comparison, we apply the same scaling argument to flows that are initially unstably segregated. The numerical results in figure~\ref{fig:5}D show that the maximum-mixing timescale, which is attained prior to the long-time segregated state (figure~\ref{fig:5}B), follows the same approximate scaling, $\hat{\tau}_m\propto Pe^{-0.6}$, over a broader range of $Pe$. In contrast to the initially stable case, however, $\hat{\tau}_m$ retains a measurable dependence on $Pe$ even at low-$Pe$. This behaviour is expected because, for an initially unstable arrangement, the early evolution is governed by segregation-driven rearrangement that first enhances interpenetration between species and temporarily increases mixing before the system relaxes towards a new stable state in which the degree of mixing decreases as segregation becomes dominant.

\section{Concluding remarks and outlook}\label{sec:6}

We have examined how segregation and mixing evolve in shear-driven bidisperse granular flows, focusing on the competition between size-driven segregation and diffusive remixing. Using the advection--segregation--diffusion continuum model \citep{Gray2018} under uniform shear, we performed systematic numerical experiments to quantify the roles of the particle-size ratio $R_d=d_l/d_s>1$ and the Péclet number $Pe$, which measures the relative strength of segregation to diffusive remixing. Specifically, we characterized: (i) the temporal evolution of the degree of segregation $\mathscr{S}{\phi}(\hat{t})$ and its long-time asymptotic value $\mathscr{S}{\phi}^{\infty}$; (ii) the thickness of the bidisperse mixing-layer $\hat{\delta}_m$, where both species coexist; and (iii) the timescale $\hat{\tau}_m$ required to reach the state of maximal mixing.

Across these metrics, we derived asymptotic scalings that depend explicitly on $Pe$ and $R_d$, providing a reduced description of segregation dynamics in terms of a small set of control parameters. These results clarify how segregation strength, interface sharpness, and transient mixing times emerge from the balance between segregation fluxes and diffusive remixing. The resulting scaling laws provide a predictive framework for estimating attainable segregation states, mixing-layer thicknesses, and associated timescales, with direct implications for the design and interpretation of laboratory experiments, discrete-element simulations, and industrial granular flows.

\textcolor{black}{Several important directions remain open:
\begin{enumerate}
\item First, the present continuum description should be tested against controlled laboratory experiments to assess its quantitative accuracy. Such tests are especially needed in regimes where diffusion may be non-Fickian, shear is spatially non-uniform, or boundary effects influence the development of the mixing layer.
\item Second, discrete-element simulations offer a direct route to evaluate the assumptions underlying the model. In particular, they can be used to test the segregation-flux closure, the form of the diffusive remixing term, and the extent to which both emerge from particle-scale dynamics, contact networks, and local rearrangements.
\item Third, extending the framework to polydisperse mixtures, density-contrasted particles, non-uniform forcing, and coupled rheology--segregation feedbacks would bring the theory closer to natural and industrial granular flows, where material properties and driving conditions are rarely uniform.
\item Fourth, particle shape remains a particularly promising frontier. Shape controls packing, mobility, force-chain organization, and local rearrangements, and therefore may strongly influence how a granular system partitions imposed mechanical energy between mixing and segregation. Establishing this connection could provide new predictive tools for controlling mixed or segregated states in geophysical, agricultural, pharmaceutical, and industrial granular materials.
\item Finally, a central unresolved problem is to determine whether a universal instability criterion exists  for the onset of mixing in stably segregated systems, and conversely, for the onset of segregation in initially mixed or stably segregated systems. Addressing this question would help identify the threshold shear stress, strain rate, or energy input required to drive a granular mixture away from equilibrium, toward remixing, or toward a new segregated equilibrium.
\end{enumerate}}

The broader challenge is therefore not only to predict when granular materials segregate, but to learn how to control when they mix.

\vskip6pt

\enlargethispage{20pt}

\dataccess{The data obtained from the numerical solutions for both unstable and stable cases can be found in the  \href{https://doi.org/10.5281/zenodo.19767321}{repository} allocated at Zenodo \cite{ZenodoRSTA}.}

\aucontribute{H.N.U: conceptualization, formal analysis, investigation, methodology, writing original draft, funding acquisition. T.T.: conceptualization, perform numerical simulations, formal analysis, investigation, methodology, writing original draft, funding acquisition.}

\funding{H.N.U.~was~supported~by~the~University of Pennsylvania~start-up grant. T.T. received support from Agencia Nacional de Investigación y Desarrollo (ANID) through FONDECYT Iniciación Project 11240630.}

\conflict{We declare we have no competing interests.}

\ack{H.N.U. thanks the Guest Editors for the invitation to submit an article to the present issue. The authors thank the two anonymous reviewers for their constructive comments and suggestions, which helped improve the manuscript.}

%%%%%%%%%% Insert bibliography here %%%%%%%%%%%%%%

\vskip2pc

\bibliographystyle{RS}
\bibliography{references}

\end{document}